\documentclass{article}
\def\vol#1#2#3{{\bf {#1}} ({#2}) {#3}}
\def\NP{Nucl.~Phys. }
\def\PL{Phys.~Lett. }
\def\PR{Phys.~Rev. }
\def\PRP{Phys.~Rep. }

\def\IJMP{Int.~J.~Mod.~Phys. }

\def\2tvec#1#2{
\left(
\begin{array}{c}
#1  \\
#2  \\
\end{array}
\right)}

\def\mat2#1#2#3#4{
\left(
\begin{array}{cc}
#1 & #2 \\
#3 & #4 \\
\end{array}
\right)
}

\def\Mat3#1#2#3#4#5#6#7#8#9{
\left(
\begin{array}{ccc}
#1 & #2 & #3 \\
#4 & #5 & #6 \\
#7 & #8 & #9 \\
\end{array}
\right)
}

\def\3tvec#1#2#3{
\left(
\begin{array}{c}
#1  \\
#2  \\
#3  \\
\end{array}
\right)}

\def\4tvec#1#2#3#4{
\left(
\begin{array}{c}
#1  \\
#2  \\
#3  \\
#4  \\
\end{array}
\right)}

\def\L{\left}
\def\R{\right}

\def\pl{\partial}

\def\hbar{\hspace{1mm}\bar{}\hspace{-1mm}h}

\def\eqn#1{
\begin{eqnarray}
#1
\end{eqnarray}
}

\begin{document}

\begin{titlepage}
\begin{center}

\vspace{-11.5cm}
\begin{flushright}
KANAZAWA-10-08
\end{flushright}

\vspace{1cm}
{\bf\Large Seesaw Mechanism Confronting PAMELA\\[1mm]
  in $S_4$ Flavor Symmetric 
Extra U(1) Model}
\vspace{1cm}

Yasuhiro Daikoku$^*$$^\ddagger$ \footnote{E-mail: yasu\_daikoku@yahoo.co.jp}, 
\quad 
Hiroshi Okada$^\dagger$\footnote{E-mail: HOkada@bue.edu.eg},
  \quad and \quad 
  Takashi Toma$^*$$^\ddagger$ \footnote{E-mail: t-toma@hep.s.kanazawa-u.ac.jp} 

\vspace{5mm}
  {\it Institute for Theoretical Physics, Kanazawa University, Kanazawa
  920-1192, Japan}$^*$$^\ddagger$ 
 
 \vspace{5mm}

   {\it Centre for Theoretical Physics, The British University in 
Egypt,\\[1mm] El Sherouk City, Postal No, 11837, P.O. Box 43, Egypt
}$^\dagger$

\vspace{8mm}

\begin{abstract}
We study cosmic-ray anomaly observed by PAMELA based on 
$E_6$ inspired extra U(1) model with $S_4$ flavor symmetry.
In our model, the lightest flavon has very long lifetime of
${\cal O}(10^{18})$ second which is longer than the age of the universe, but not long
 enough to explain the PAMELA result ${\cal O}(10^{26})$ sec.
 Such a situation could be avoidable by considering
 that the flavon is not the dominant component of dark matters.
 However non-thermalizing the flavon is needed to obtain proper relic density.
This relates reheating temperature of the universe with seesaw mass scale.
If we assume this flavon is a particle decaying into positron (or electron),
the seesaw mass scale is constrained by reheating temperature.
Thus we find an interesting result that the allowed region is around
${\cal O}(10^{12})\ \rm{GeV}$, which is consistent with our original result.
\end{abstract}

\end{center}
\end{titlepage}

\setcounter{footnote}{0}

\section{Introduction}


It is one of the important task to build more economical models in (non-)Abelian flavor symmetries.
In such the framework of models with non-renormalizable operators especially, 
couplings of the terms are usually suppressed by high energy cut-off scale.
Therefore gauge singlet neutral bosons 
which couple to the term may play an important role
not only to construct the mass matrix forms but also to be promising dark matter candidates,
because they could be less-interactive enough.
It is recently known that the dark matter can be a good candidate
to explain PAMELA data \cite{pamela}.
Subsequently,
there are many attempts to explain the positron anomaly 
by annihilation \cite{annihilation} or decay of the dark matter \cite{positron}.
According to constraint from diffuse gamma ray \cite{gamma},
an interpretation by annihilation is almost excluded.
Thus the PAMELA result is in favor of the decaying dark matter,
when it has the life time of $\Gamma^{-1}\sim {\cal O}(10^{26})$ sec.
This is much longer than the age of the universe.

In this paper,
we study such a cosmic-ray excess by a singlet scalar (flavon)
in $S_4$ flavor model \footnote{It requires exotic scalar quarks, which induce proton decay.
In our original work \cite{s4e6}, we have shown that
proton decay  is suppressed by $S_4$ flavor symmetry very well.}
of a supersymmetric \cite{SUSY} with $E_6$ inspired extra U(1) gauge symmetry \cite{mu-problem}. 
The flavor symmetry is broken by vacuum expectation value (VEV) 
of the flavon and this VEV gives large Majorana masses for right handed 
neutrinos (RHNs). As the flavon couples to standard model particles only through 
non-renormalizable operators,
the life time of the flavon could be longer than the age of the universe.
However the life time of the flavon in our model is not longer than ${\rm O}(10^{26})$ sec. 
Therefore we consider that the flavon is not the dominant component
of dark matters. Then the most leading interaction which the flavon
has is extra U(1)$_Z$ interaction and the interaction is extremely weak
because it is suppressed by the large mass scale of the U(1)$_Z$ gauge boson.
As the result, annihilation cross section is too small to obtain proper relic density of the flavon
as long as we assume that the flavon is in thermal equilibrium. 
So we consider that the flavon is never in thermal equilibrium.
Non-thermalizing the flavon constrains reheating temperature and also relates 
to right-handed neutrino mass scale.
If we assume reheating temperature is constrained as 
$10^7\ {\rm GeV} > T_{{\rm RH}} > 10^4\ {\rm GeV}$,
right-handed neutrino mass scale should be around $10^{12}\ {\rm GeV}$, 
which is consistent with our original result.

The paper is organized as follows.
In section 2, we review the basic structure of $S_4$ flavor symmetric 
extra U(1) model.
We evaluate density parameter of lightest flavon from PAMELA constraint in section 3,
and estimate required right-handed neutrino mass scale for explaining it in 
section 4. Finally we make a brief summary in section 5.

\section{The Extra U(1) Model with $S_4$ Flavor Symmetry}

\subsection{The Extra U(1) Model}

The basic structure of the extra U(1) model is given as follows 
\cite{s4e6}.
At high energy scale, the gauge symmetry of model has two extra U(1)s,
which consists maximal subgroup of $E_6$ as $G_2=G_{SM}\times U(1)_X\times 
U(1)_Z\subset E_6$.
MSSM superfields and additional superfields are embedded in three 27 
multiplets of $E_6$ to cancel anomalies,
as ${\bf 27}\supset \L\{Q,U^c,E^c,D^c,L,N^c,H^D,g^c,H^U,g,S\R\}$,
where $N^c$ are right-handed neutrinos (RHN), $g$ and $g^c$ are  exotic 
quarks (g-quark), and $S$ are  $G_{SM}$ singlets, which is illustrated in 
Table 1.
We introduce $G_{SM}\times U(1)_X$ singlets $\Phi$ and $\Phi^c$ 
which develop the intermediate scale VEVs along the D-flat direction
of $\L<\Phi\R>=\L<\Phi^c\R>$, then the $U(1)_Z$ is broken and the RHNs 
obtain the mass terms. 
After the symmetry is broken, as the R-parity symmetry
remains unbroken, $G_1=G_{SM}\times U(1)_X\times R$ survives at low energy.
This is the symmetry of the low energy extra U(1) model.

Within the renormalizable operators, $G_2$ symmetric superpotential is 
given as follows:
\eqn{
W_2&=&W_0+W_S+W_B, \\
W_0&=&Y^UH^UQU^c+Y^DH^DQD^c+Y^EH^DLE^c+Y^N H^ULN^c+Y^M\Phi N^cN^c, \\
W_S&=&kSgg^c+\lambda SH^UH^D, \\
W_B&=&\lambda_1 QQg+\lambda_2 g^cU^cD^c+\lambda_3 gE^cU^c+\lambda_4 
g^cLQ+\lambda_5gD^cN^c.
}
Where $W_0$ is the same as the superpotential of the MSSM with the RHNs 
besides the absence of $\mu$-term, and
$W_S$ and $W_B$ are the new interactions.
In $W_S$, $kSgg^c$ drives the soft SUSY breaking
scalar squared mass of S to negative through the renormalization group 
equations (RGEs) and
then breaks $U(1)_X$ and generates mass terms of g-quarks, and $\lambda 
SH^UH^D$ is source of the effective $\mu$-term.
Therefore, $W_0$ and $W_S$ are phenomenologically necessary.
In contrast, $W_B$ 
leads to very rapid proton decay and  must be forbidden. 
This is done by $S_4$ flavor symmetry.

\begin{table}[htbp]
\begin{center}
\begin{tabular}{|c|c|c|c|c|c|c|c|c|c|c|c||c|c|}
\hline
         &$Q$ &$U^c$    &$E^c$&$D^c$    &$L$ &$N^c$&$H^D$&$g^c$    
&$H^U$&$g$ &$S$ &$\Phi$&$\Phi^c$\\ \hline
$SU(3)_c$&$3$ &$3^*$    &$1$  &$3^*$    &$1$ &$1$  &$1$  &$3^*$    &$1$  
&$3$ &$1$ &$1$   &$1$     \\ \hline
$SU(2)_W$&$2$ &$1$      &$1$  &$1$      &$2$ &$1$  &$2$  &$1$      &$2$  
&$1$ &$1$ &$1$   &$1$     \\ \hline
$y=6Y$   &$1$ &$-4$     &$6$  &$2$      &$-3$&$0$  &$-3$ &$2$      &$3$  
&$-2$&$0$ &$0$   &$0$     \\ \hline
$x$      &$1$ &$1$      &$1$  &$2$      &$2$ &$0$  &$-3$ &$-3$     &$-2$ 
&$-2$&$5$ &$0$   &$0$     \\ \hline
$z$      &$-1$&$-1$     &$-1$ &$2$      &$2$ &$-4$ &$-1$ &$-1$     &$2$  
&$2$ &$-1$&$8$   &$-8$    \\ \hline
$R$      &$-$ &$-$      &$-$  &$-$      &$-$ &$-$  &$+$  &$+$      &$+$  
&$+$ &$+$ &$+$   &$+$     \\ \hline
\end{tabular}
\end{center}
\caption{$G_2$ assignment of fields.
Where the $x$, $y$ and $z$ are charges of $U(1)_X$, $U(1)_Y$ and $U(1)_Z$, 
and $Y$ is hypercharge.}
\end{table}

\subsection{$S_4$ Flavor Symmetry}

Non-Abelian group $S_4$ has two
singlet representations ${\bf 1}$, ${\bf 1'}$, one doublet
representation ${\bf 2}$ and two triplet representations ${\bf
3}$, ${\bf 3'}$, where ${\bf 1}$ is the trivial representation \cite{s4}. 
The most essential structure of $S_4$ group is that multiplication of two 
doublets
does not contain triplets.
With this property, 
if $g$ and $g^c$ are assigned to triplets and the others are assigned to 
singlets
or doublets, then $W_B$ is forbidden. 


The absence of $W_B$ makes g-quarks and proton stable, but
the existence of g-quarks which have life time longer than
0.1 second spoils the success of Big Ban nucleosynthesis. In order
to evade this problem, 
we assign $\Phi^c$ as triplet of $S_4$ and add the non-renormalizable 
terms: \eqn{
W_{NRB}=\frac{1}{M^2_P}\Phi\Phi^c\L(QQg+g^cU^cD^c+gE^cU^c+g^cLQ+gD^cN^c\R). 
\label{nrb}
}
When $\Phi^c$ develops VEV with 
\eqn{
\frac{\L<\Phi\Phi^c\R>}{M^2_P}\sim 10^{-12},    \label{condition} } 
the phenomenological constraints on the life times of proton and
g-quarks are satisfied at the same time \cite{f-extra-u1},
and the right-handed neutrino mass scale can be predicted as 
$M_R\sim \langle\Phi\rangle\sim 10^{-6}M_P\sim 10^{12}\ {\rm GeV}$.


As the flavons $\Phi$ and $\Phi^c$ which are the triggers of flavor 
violation do not
have renormalizable interactions with light particles, the lightest flavon 
(${\cal LF}$)
has very long life time. In following sections, 
we consider whether this particle explains PAMELA observation.

\section{Flavon Decay Width}
With the assignment that $\Phi^c$ is $S_4$ triplet and $\Phi$ is singlet or 
doublet,
the leading term of flavon superpotential is given by
\eqn{
W_\Phi=\frac{a}{M_P}\Phi^2(\Phi^c)^2.
}
Solving  the potential minimum conditions, we get
\eqn{
V=|\Phi|=|\Phi^c|\sim (m_{SUSY}M_P/a)^\frac12\sim 
10^{11}a^{-\frac12}\L(\frac{m_{SUSY}}{10\ \rm{TeV}}\R)^\frac12\ \rm{GeV},
}
where $a\sim 10^{-2}$ is required from Eq.(\ref{condition}). 
Integrating out the heavy RHNs, we get effective seesaw operators as 
follows
\eqn{
W_{eff}=\frac{1}{Y^M\Phi}(Y^NH^UL)^2. \label{seesaw}
}
Here we redefine flavon as perturbation around VEV as $\Phi\to V+\Phi$, 
then Eq.(\ref{seesaw}) 
is rewritten as
\eqn{
W_{eff}=\frac{1}{Y^MV}(Y^NH^UL)^2-\frac{\Phi}{Y^MV^2}(Y^NH^UL)^2.
\label{perseesaw}
}
If sleptons, squarks, g-quarks and scalar g-quarks are heavier than ${\cal LF}$, 
this operator gives dominant contribution 
to decay width of ${\cal LF}$ through
\eqn{
{\cal L}_{eff}=\frac{m_\nu}{V}\Phi\nu\nu,
\label{dw-LF}
}
where we assume
\eqn{
Y^M\sim 1,\quad m_\nu\sim \frac{(Y^Nv)^2}{V},\quad \L<H^U\R>= 
\frac{v}{\sqrt{2}}.
}
From the interaction, monochromatic intense diffuse neutrino flux is
expected in cosmic-ray at $E_\nu=m_\Phi/2$. It is an important signature of
the decaying dark matter model.
Considering the superpotential $W_\Phi$,
it is easily  shown that only one linear combination of six flavons 
$\Phi_{1,2,3},\Phi^c_{1,2,3}$ has super heavy mass 
around $V$ and another five flavons have ${\cal O}(M_{SUSY})$ masses. As is 
pointed in ref. \cite{nupamela}, 
this interaction is good candidate for explaining PAMELA phenomena, because 
if $V$ is around  $10^{16}\ {\rm GeV}$,
then the partial decay width of $\Phi\to \nu H^-e^+$ is given by
\eqn{
\Gamma^{-1}(\Phi\to \nu 
H^-e^+)=\L(\frac{m^3_\Phi}{768\pi^3v^2}\frac{m^2_\nu}{V^2}\R)^{-1}=3.5\times 10^{26}\ {\rm sec},
\label{dw} }
where
\eqn{
m_\nu=0.1\ {\rm eV},\quad v=246\ {\rm GeV},\quad m_\Phi=3\ {\rm TeV},\quad V=10^{16}\ {\rm GeV}.
}
The result of Eq.(\ref{dw}) 
is in good agreement with ref. \cite{positron}.
However, as $V\sim 10^{12}\ {\rm GeV}$ in our model, the life time of ${\cal LF}$ is not 
$10^{26}\ {\rm sec}$ but $10^{18}\ {\rm sec}$.
So we assume  this ${\cal LF}$ is not the dominant component of dark matter 
($\Omega_{LF}\ll\Omega_{DM}$).
Introducing mixing parameter $\epsilon$ defined as
\eqn{
\Phi=\epsilon \Phi_{LF}+\cdots,
}
where $\Phi_{ LF}$ is the lightest flavon field and 
the dots $\cdots$ means contributions from heavier  flavons, 
the partial decay width of $\Phi_{LF}\to\nu H^-e^+$ is given by
\eqn{
\Gamma^{-1}(\Phi_{LF}\to\nu H^-e^+)=3.5\times 10^{18}\epsilon^{-2}
\L(\frac{V}{10^{12}\ {\rm GeV}}\R)^2\ {\rm sec} .
}
In order to explain  positron  flux observed by PAMELA, density parameter 
of ${\cal LF}$ should be
\eqn{
\Omega_{LF}=\frac{\Gamma^{-1}(\Phi_{LF}\to\nu 
H^-e^+)}{10^{26}\ {\rm sec}}\Omega_{DM}
=3.9\times 10^{-9}\epsilon^{-2}\L(\frac{V}{10^{12}\ {\rm GeV}}\R)^2/h^2,
\label{lf-relic}
}
where we use the observed value of density parameter of dark matter in \cite{PDG2008} as
\eqn{\Omega_{DM}=0.11/h^2.}

Before calculating density parameter of ${\cal LF}$ in next section, we give some 
comments about possible decay channels of flavons.
The interactions between flavons are described by
\eqn{
\L(\frac{a}{M_P}\R)^2V^3\Phi_i\Phi_j\Phi_k\sim 
(10^{-4}\ {\rm GeV})\Phi_i\Phi_j\Phi_k,
}
from which decay width of heavier flavon to lighter flavons is given by
\eqn{
\Gamma(\Phi_i\to\Phi_j\Phi_k)\sim 10^{12}\ {\rm sec}^{-1}.
}
With this interactions, all heavier flavons decay into ${\cal LF}$ finally.
If ${\cal LF}$ is heavier than sneutrinos,  the following operators coming from Eq.(\ref{perseesaw}) 
 may contribute  to ${\cal LF}$ decay:
\eqn{
{\cal L}_{\Phi NN}\sim m_{SUSY}\frac{m_\nu}{V}\Phi_{LF}NN.
}
Where $N$ is sneutrino. 
If ${\cal LF}$ is heavier than scalar g-quarks, the following operators coming from Eq.(\ref{nrb}) 
may contribute too: 
\eqn{\frac{V}{M^2_P}\Phi_{LF}gqq,}
where $g$ is scalar g-quark and q is doublet quark. 
The contributions from these operators are the same order as the contribution 
from Eq.(\ref{dw-LF}). 
Hereafter, we assume ${\cal LF}$ is lighter than those scalar particles and 
g-quarks.  
In this case, ${\cal LF}$ decay channels to these particles
are kinematically closed. 

\section{Estimation of RHN Mass Scale}

At high temperature $T\gg m_{SUSY}$, dominant contribution to
$\chi\chi\to\Phi\Phi$ come from U(1)$_Z$ gauge interaction,
where $\chi$ is the particle in thermal bath.
This interaction is in thermal equilibrium  when
\eqn{\Gamma(\chi\chi\to\Phi\Phi)\sim \frac{T^5}{V^4}> H\sim \frac{T^2}{m_P}
\label{thermal-eq} }
is satisfied. Note that $M_P=2.43\times 10^{18}\ {\rm GeV}$ is reduced Planck mass 
and $m_P=1.22\times 10^{19}\ {\rm GeV}$ is Planck mass.
Evading overproduction of gravitino, $T<10^7\ {\rm GeV}$ must be satisfied with
\cite{gravitino}. Then the constraint Eq.(\ref{thermal-eq}) 
leads following condition:
\eqn{10^{10}\ {\rm GeV}>V.}
However it is difficult to satisfy this condition in our model.
Moreover, if flavon is in thermal equilibrium,  there is a problem of 
over-production of flavon
because U(1)$_Z$ gauge interaction is too weak at low temperature
to cause appropriate pair annihilation of flavons.
Based on this discussion, we assume U(1)$_Z$ gauge interaction is never in 
thermal equilibrium and the universe starts 
with low flavon number density $n_\Phi(T_{RH})=0$.

The interaction between U(1)$_Z$ gauge boson $A_\mu$ and chiral multiplet 
$(\psi,\Psi)$,
where $\psi$ is fermion and $\Psi$ is boson, is given by
\eqn{
{\cal L}_{gauge}=ig_zA^\mu\sum_iz_i\L[\bar{\psi}_{i,L}\gamma_{\mu} 
\psi_{i,L}
+\Psi_i\pl_\mu\Psi^\dagger_i-\Psi^\dagger_i\pl_\mu\Psi_i\R],
}
from which thermally averaged  cross sections are given by
\eqn{
\sum_\psi\L<\sigma_{\psi\psi\to\phi\phi}|v|\R>n^2_\psi&=&55.1CT^8, \\
\sum_\Psi\L<\sigma_{\Psi\Psi\to\phi\phi}|v|\R>n^2_\Psi&=&96.1CT^8, \\
\sum_\psi\L<\sigma_{\psi\psi\to\Phi\Phi}|v|\R>n^2_\psi&=&96.1CT^8, \\
\sum_\Psi\L<\sigma_{\Psi\Psi\to\Phi\Phi}|v|\R>n^2_\Psi&=&248.8CT^8, \\
C&=&\frac{21}{(2\pi)^5}\L(\frac{g^2_zz_\Phi}{M^2_g}\R)^2, \\
M_g&=&2g_zz_\Phi V=16g_zV,
}
where $\sum_{\psi,\Psi}$ does not contain flavon $\Phi$ and its fermion 
partner $\phi$
because $n_\Phi$ and $n_\phi$ are negligible.
Here we define the ${\cal LF}$ production rate $N_{LF}$ which means how many ${\cal LF}$s are 
generated 
per one degree of freedom of flavon and its fermion partner.
Because the total degrees of freedom is 20, $N_{LF}$ is bounded from below 
by $1/20$
\footnote{Note that 4 of $24=4\times6$ are super-heavy or eaten by gauge boson, 
where $6$ is total number of flavon superfields $(\Phi,\Phi^c)$.}.
If we require that the masses of flavons and those fermion partners are 
smaller than $9\ {\rm TeV}$,
$N_{LF}$ is bounded from above by $9\ {\rm TeV}/m_{LF}=3$.
Therefore allowed range of $N_{LF}$ is given by
\eqn{ 3\geq N_{LF} \geq 0.05.\label{nfl} }
The model dependence of flavon sector affects this analysis only through 
$\epsilon$ and $N_{LF}$.
Using $N_{LF}$, the Boltzmann equation for the ${\cal LF}$ number density $n_{LF}$:
\eqn{
n_{LF}=5N_{LF}(n_\phi+n_\Phi),
}
is given by
\eqn{
\dot{n}_{LF}+3Hn_{LF}=2480N_{LF}CT^8,
\label{boltz-eq}
}
where Hubble constant $H$ is given by
\eqn{
H&=&1.66\sqrt{g_*}\frac{T^2}{m_P}, \\
g_*&=&341.25.
}
Solving Eq.(\ref{boltz-eq}) 
with boundary condition $n_{LF}(T_{RH})=0$, ${\cal LF}$ to entropy 
ratio is calculated as follow:
\eqn{
\L(\frac{n_{LF}}{s}\R)_0=
\frac{15\times 2480\times 21 m_PN_{LF}}{2\pi^2g_*\times 
30.67(2\pi)^5}\L(\frac{g^2_zz_\Phi}{M^2_g}\R)^2
T^3_{RH}.\label{entropy-ratio}
}
Note that Eq.(\ref{entropy-ratio}) 
does not depend on the definition of
$U(1)_Z$ gauge coupling constant $g_z$ and charge normalization.
Finally  we require the density parameter
\eqn{
\Omega_{LF}=\frac{m_{LF}n_{LF}}{\rho_c}
}
satisfies Eq.(\ref{lf-relic}) 
then we get
\eqn{
\frac{V}{10^{12}\ {\rm GeV}}=\L(\epsilon^2N_{LF}\R)^\frac16
\L(\frac{T_{RH}}{10^5\ {\rm GeV}}\R)^\frac12,
}
where
\eqn{
s_0&=&2890/{\rm cm}^3 , \\
\rho_c&=&1.05\times 10^4 h^2\ {\rm eV}/{\rm cm}^3,
}
are used \cite{PDG2008}.

If we assume 
\eqn{
10^7\ {\rm GeV}\geq T_{RH} \geq 10^4\ {\rm GeV} \gg m_{LF},
}
and the model dependent parameters as Eq.(\ref{nfl}) 
and
\eqn{
1\geq \epsilon \geq 0.1,
}
we get the allowed range of $V$ as follow:
\eqn{
12\geq \frac{V}{10^{12}\ {\rm GeV}} \geq 0.09.
}
This prediction is consistent with our previous result $V\sim 10^{12}\ {\rm GeV}$ 
which comes from phenomenological 
constraints for proton life time and g-quark life time \cite{s4e6}.
If we use $V=10^{12}\ {\rm GeV}$, reheating temperature is predicted as
\eqn{
T_{RH}=(0.7-5.8)\times 10^5\ {\rm GeV}.
}

Finally we give some comments.
In the case that ${\cal LF}$ is dominant component of dark matter as 
$\Omega_{LF}\sim \Omega_{DM}$,
which is realized by putting  $\epsilon(10^{12}\ {\rm GeV}/V)\sim 2\times 10^{-4}$, 
$\epsilon$ is bounded from above by
\eqn{
\epsilon<1.2\times 10^{-4}.
}
However, it seems difficult to explain such a small mixing angle. 

If the final states of ${\cal LF}$ decay contain quarks, anti-proton flux may be
significantly induced, which conflicts with cosmic-ray observations \cite{Adriani:2008zq}.
This problem is avoided when ${\cal LF}$ decay mainly into charged Higgs which
does not couple to quarks. Such a type of model \cite{Haba:2010ag}
can be constructed with $S_4$ flavor
symmetry. For example, if all quarks are assigned to $S_4$-singlets and
$SU(2)_W$ Higgs doublets $H^U$ and $H^D$ are assigned to 
one $S_4$-doublet and one $S_4$-singlet respectively,
Yukawa interactions between $S_4$-doublet Higgs and quarks are forbidden.
This assignment also solves Higgs-FCNC problem, as pointed in ref. 
\cite{FCNC}.
However, as there is no room for model building in weak interaction,
we can not suppress anti-proton production by weak interaction such as
$\Phi\rightarrow \nu W^-e^+\rightarrow\bar{p}$.
It is not obvious whether the effects of weak interaction spoil this 
solution or not,
which is left  for future work.


\section{Summary}

In this paper, we have considered cosmic-ray anomaly observed by
PAMELA based on $S_4$ flavor symmetric extra U(1) model.
Identifying the particle decaying into positron as the lightest flavon
which is the sub-dominant component of dark matter, reheating temperature
of the universe and the mass scale of right handed neutrinos were
related, and
we estimated the mass scale of right handed neutrinos from the relation.
As a result, the constraint that reheating temperature is low enough to 
suppress gravitino production
gives vacuum expectation value of flavon around $10^{12}\ {\rm GeV}$, which is 
consistent with the prediction of our model.
This result supports the idea of neutrino seesaw mechanism based on heavy 
right handed neutrino with $O(10^{12})\ {\rm GeV}$ mass.
Finally, monochromatic diffuse neutrino flux is expected at $E_\nu=m_\Phi/2$ in cosmic-ray as a predction of the model.

\section*{Acknowledgments}
H.O. thanks great hospitality of the Kanazawa Institute for
Theoretical Physics at Kanazawa University. Discussions during my
visit were fruitful to finalize this project.
H. O. acknowledges partial support from the Science and Technology 
Development Fund (STDF) project ID 437 and the ICTP project ID 30.
The numerical calculations were carried out on Altix3700 BX2 at YITP in
Kyoto University.




\begin{thebibliography}{99}

\bibitem{pamela}O.~Adriani {\em et al}. [PAMELA Collaboration], Nature {\bf 
458} (2009) 607.

\bibitem{annihilation}M. Beltran, D. Hooper, E. W. Kolb and
	Z. A. C. Krusberg, Phys. Rev. {\bf D80} (2009) 043509;
	M. Cirelli, M. Kadastik, M. Raidal and A. Strumia,
	Nucl. Phys. {\bf B813} (2009) 1; 
	J. Hisano, S. Matsumoto and M. M. Nojiri,
	Phys. Rev. Lett. {\bf 92} (2004) 031303;
	D. Feldman, Z. Liu and P. Nath,
Phys. Rev. {\bf D79} (2009) 063506; 
	M. Ibe, H. Murayama, T. T. Yanagida,
	Phys. Rev. {\bf D79} (2009) 095009; 
	 A.~A.~El-Zant, S.~Khalil and H.~Okada,
  Phys.\ Rev.\  D {\bf 81}, 123507 (2010);
	D. Suematsu, T. Toma and T. Yoshida,
	Phys. Rev. {\bf D82} (2010) 013012.

\bibitem{positron}
P.~Meade, M.~Papucci, A.~Strumia and T.~Volansky, \NP\vol{B831}{2010}{178};
A.~Ibarra, D.~Tran and C.~Weniger, JCAP\vol{1001}{2010}{009};
C.~R.~Chen, S.~K.~Mandal and F.~Takahashi, JCAP\vol{1001}{2010}{023};
J.~Liu, Q.~Yuan, X.~Bi, H.~Li and X.~Zhang, arXiv:0911.1002;
M.~Cirelli, P.~Panci and P.~D.~Serpico, arXiv:0912.0663;
G.~Hutsi, A.~Hektor and M.~Raidal, arXiv:1004.2036;
A. Ibarra and D. Tran, JCAP\vol{0807}{2008}{002};
A. Ibarra and D. Tran, JCAP\vol{0902}{2009}{021}.

	
\bibitem{gamma}G. Bertone, M. Cirelli, A. Strumia and M. Taoso,
	JCAP {\bf 03} (2009) 009; M. Cirelli and
	P. Panci, Nucl. Phys. {\bf B821}, (2009) 399; M. Papucci and
	A. Strumia, JCAP {\bf 03} (2010) 014.
	

\bibitem{SUSY}H.P. Nilles,  \PRP\vol{110}{1984}{1}.

\bibitem{mu-problem}
D. Suematsu and Y. Yamagishi, \IJMP\vol{A10}{1995}{4521}.

\bibitem{s4e6}
  Y.~Daikoku and H.~Okada,
  Phys.\ Rev.\  D {\bf 82}, 033007 (2010).


\bibitem{s4}
S. Pakvasa and H. Sugawara,  \PL\vol{B73}{1978}{61};
E. Ma, \PL\vol{B632}{2006}{352}; 
C. Hagedorn, M. Lindner and R.N. Mohapatra, JHEP\vol{0606}{2006}{042}; 
 H.~Zhang,
  Phys.\ Lett.\  B {\bf 655}, 132 (2007);
Y. Koide, JHEP\vol{0708}{2007}{086};
F. Bazzocchi and S. Morisi, 
Phys.\ Rev.\  D {\bf 80}, 096005 (2009);
H.~Ishimori, Y.~Shimizu and M.~Tanimoto,
  Prog.\ Theor.\ Phys.\  {\bf 121}, 769 (2009);
F.~Bazzocchi, L.~Merlo and S.~Morisi,
  Nucl.\ Phys.\  B {\bf 816} (2009) 204;
F.~Bazzocchi, L.~Merlo and S.~Morisi,
  Phys.\ Rev.\  D {\bf 80} (2009) 053003
   G.~Altarelli, F.~Feruglio and L.~Merlo,
  JHEP {\bf 0905} (2009) 020;
  W.~Grimus, L.~Lavoura and P.~O.~Ludl,
  J.\ Phys.\ G {\bf 36}, 115007 (2009);
  G.~J.~Ding,
  Nucl.\ Phys.\  B {\bf 827}, 82 (2010);
   B.~Dutta, Y.~Mimura and R.~N.~Mohapatra,
   JHEP {\bf 1005}, 034 (2010);
  D.~Meloni,
  J.\ Phys.\ G {\bf 37}, 055201 (2010); 
  S.~Morisi and E.~Peinado,
  Phys.\ Rev.\  D {\bf 81}, 085015 (2010);
    G.~Altarelli and F.~Feruglio,
  arXiv:1002.0211 [hep-ph];
  H.~Ishimori, T.~Kobayashi, H.~Ohki, H.~Okada, Y.~Shimizu and M.~Tanimoto,
  Prog.\ Theor.\ Phys.\ Suppl.\  {\bf 183}, 1 (2010);
 C.~Hagedorn, S.~F.~King and C.~Luhn,
  JHEP {\bf 1006}, 048 (2010).


\bibitem{f-extra-u1}R. Howl and S.F. King, JHEP\vol{0805}{2008}{008}. 

\bibitem{nupamela}
  S.~Matsumoto and K.~Yoshioka,
  Phys.\ Rev.\  D {\bf 82}, 053009 (2010).

\bibitem{PDG2008}C. Amsler et al. (Particle Data Group), Phys. Lett. B667, 
1 (2008) and
2009 partial update for the 2010 edition.
Cut-off date for this update was January 15, 2009.

\bibitem{gravitino}
M.~Kawasaki, K.~Kohri and T.~Moroi, 
\PR\vol{D71}{2005}{083502}[astro-ph/0408426].

	
\bibitem{Adriani:2008zq}
O.~Adriani {\it et al.},
  Phys.\ Rev.\ Lett.\ {\bf 102} (2009) 051101.


\bibitem{Haba:2010ag}
An $A_4$ symemtric dark matter is considered to explain PAMELA data in the follwoing paper.
We have also checked other non-Abelian discrete flavor symmetries such as $S_4$ can be adaptable for the attempt.\\    
N.~Haba, Y.~Kajiyama, S.~Matsumoto, H.~Okada and K.~Yoshioka,
  arXiv:1008.4777 [hep-ph].\\
For the other applications of dark matters in non-Abelian discrete symmetries; see for example,  
   M.~Hirsch, S.~Morisi, E.~Peinado and J.~W.~F.~Valle,
  arXiv:1007.0871 [hep-ph].
    
\bibitem{FCNC}Y.~Daikoku and H.~Okada, arXiv:1008.0914 [hep-ph].  

\end{thebibliography}
\end{document}